\documentclass[doublecol]{epl2} 

\newcommand\emc{E=mc^{2}}

\usepackage{amsmath}
\usepackage{amsfonts}
\usepackage{amssymb}
\usepackage{graphicx}
\usepackage{braket}
\usepackage[normalem]{ulem}

\title{Lax dynamics}
\shorttitle{Lax dynamics}
\author{Stefano Lepri \inst{1,2}}
\shortauthor{S. Lepri}
\institute{\inst{1} Consiglio Nazionale delle Ricerche, Istituto dei Sistemi Complessi, via Madonna del Piano 10, I-50019 Sesto Fiorentino, Italy\\
\inst{2} Istituto Nazionale di Fisica Nucleare, Sezione di Firenze, via G. Sansone 1, I-50019 Sesto Fiorentino, Italy}

\abstract{
A novel approach is proposed to characterize the dynamics of perturbed many-body integrable systems. 
Focusing on the paradigmatic case of the Toda chain under non-integrable 
Hamiltonian perturbations, this study introduces a method based the time evolution of the Lax eigenvalues $\lambda_\alpha$
as a proxy of the quasi-particles velocities and of the perturbed Toda actions.
A set of exact equations of motion for the $\lambda_\alpha$ is derived that 
closely resemble those for eigenenergies of a quantum problem (also known
as the Pechukas-Yukawa gas). Numerical simulations suggest that the 
invariant measure of such dynamics is basically the thermal density of states 
of the Toda lattice, regardless of the form of the perturbation. 
}

\begin{document}

\maketitle

\section{Introduction} The solution of a physical problem usually proceeds by identifying a solvable part and studying the effects of perturbations. For nonlinear systems where the solvable part is described by an integrable classical or quantum Hamiltonian, one can, with varying degrees of mathematical difficulty, separate the independent degrees of freedom (the quasiparticles) and analyze their interactions using, for example, perturbation theory. In many-body problems, 
a perturbation typically destroys integrability, leaving only a few residual conserved quantities, it is important to assess how and when thermalization, chaotic dynamics, and conventional hydrodynamic behavior occur
\cite{bastianello2021hydrodynamics}. Considering that  
a variety of physical systems as ultracold atoms, one-dimensional magnets, or optical
beams are proximate to nonlinear integrable limits \cite{doyon2025generalized}, such questions are of wide interests in many diverse contexts.

In the classical realm,  a paradigmatic example is the 
celebrated Toda lattice defined by the Hamiltonian \cite{toda2012theory} 
\begin{equation}
H_\mathrm{Toda} = \sum_{j=1}^N\big(\tfrac{p_j^2}{2} + \mathrm{e}^{-(q_{j+1} - q_j)}\big),
\label{ham}
\end{equation}
where $(q_j,p_j)$ are position and momentum of the $j$-th particle
( note that the model has no free parameter).  The discovery of its
full integrability 
\cite{henon1974integrals,Flaschka1974} sparkled a vivid 
reseach activity. 
However, its thermodynamics received only sporadic consideration \cite{theodorakopoulos1984finite,cuccoli1994thermodynamics}, and has only recently garnered renewed attention due to the formulation of Generalized Gibbs Ensembles (GGE), which extend the canonical state of standard statistical mechanics to integrable models
\cite{doyon2019generalized,spohn2020generalized,spohn2021hydrodynamic,baldovin2021statistical}.
This represent a great novelty with respect to the very many works 
dealing with zero-temperature properties  
of specific solutions e.g. solitons, breathers and nonlinear waves \cite{dauxois2006physics}, 
the emphasis being shifted to e.g. dynamical correlations at equilibrium
\cite{kundu2016equilibrium,mazzuca2023equilibrium}.

Beyond this,  insights on the effect of integrability-breaking perturbations are 
relevant for nonequilibrium properties, ranging from the classic
thermalization problem, \textit{a l\'a} Fermi-Pasta-Ulam-Tsingou (FPUT)
\cite{gallavotti2007fermi} to heat transport close to quasi-integrable limit  
\cite{benenti2023non}.
The crucial observation \cite{benettin2013fermi} is that,
$H_\mathrm{Toda}$ is the closest integrable approximation of a \textit{whole family}
of anharmonic chains with Hamiltonian of the form $H=\sum_j[ \frac12 p_j^2+
\Phi(q_{j+1} - q_j)]$.  This viewpoint has been established only 
recently \cite{benettin2021understanding},  and implies that
for a broad class of inter-particle potentials $\Phi$,    
Toda's model is a more accurate and insightful approximation 
than the standard harmonic one.  
For instance, the slow chaotic motion leading to equipartition
is not so much due to the fact a models like FPUT is a discretization of
an integrable wave equation, but rather to the fact that 
the dynamics is
(at least at low enough energies) essentially indistinguishable
from Toda's on very long time scales \cite{benettin2021understanding}.
This idea is corroborated by
numerical experiment on thermalization \cite{benettin2013fermi,goldfriend2019equilibration,benettin2023role} and stationary transport \cite{chen2014nonintegrability,DiCintio2018,dhar2019transport,lepri2020too,benenti2023non}.
In the first case, the metastable state can be seen as a
particular GGE state that slowly relaxes to standard thermal equilibrium
\cite{goldfriend2019equilibration}.

In this Letter,  
a novel point of view of the problem is presented 
based on the extension of the concept of \textit{Lax eigenvalues}, which are 
well-known in the theory of integrable models \cite{babelon2003introduction}, 
to their perturbed versions. 
It will be argued that they are insightful quantities and that 
their time-evolution caused by the perturbation has some generic
features of considerable interest. 
  
\begin{figure*}
\includegraphics[width=\textwidth]{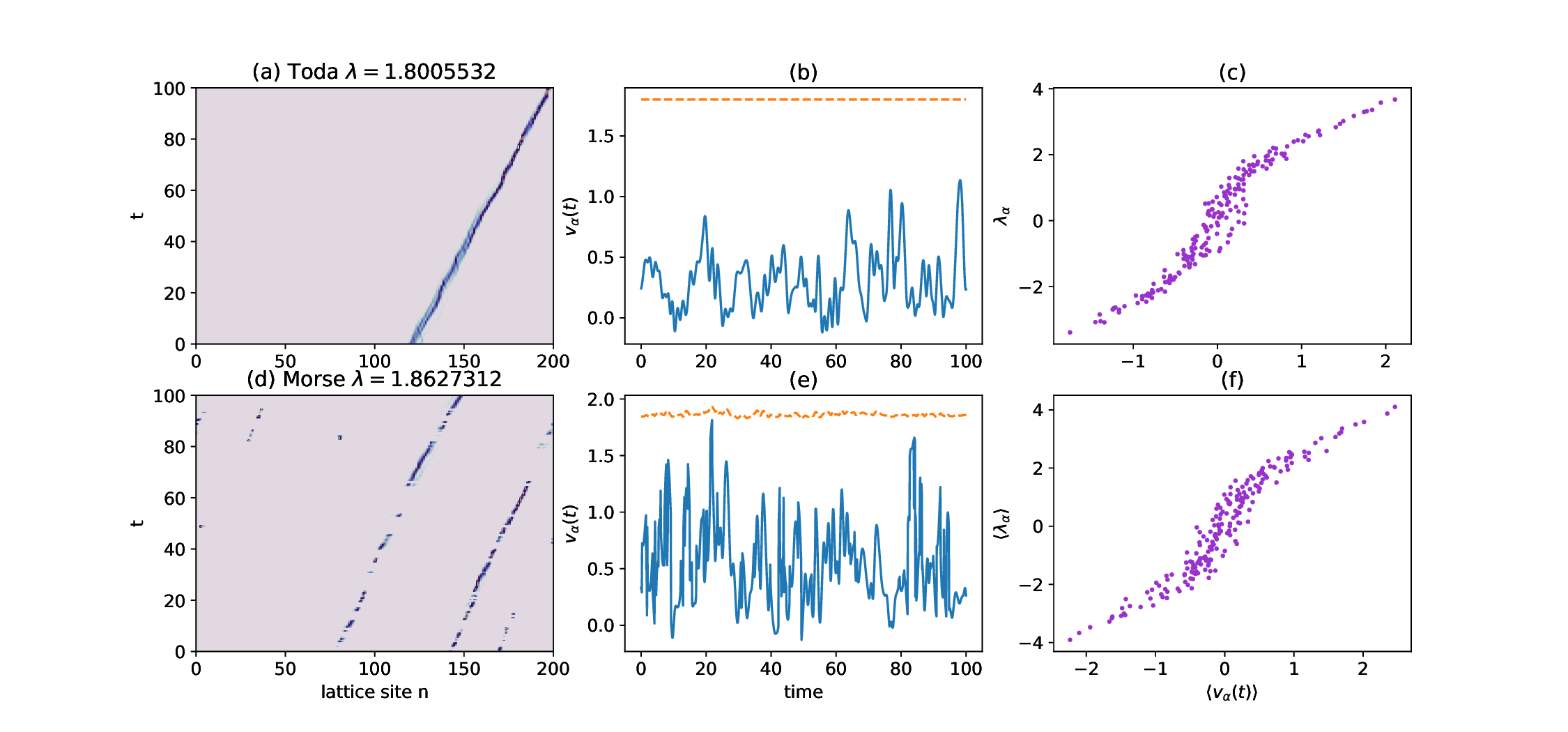}   
\caption{Simulations of Toda (upper panels) and Morse 
chains with $\varepsilon=0.1$ (lower panels); (a,d): Space-time evolution of 
the square modulus of a Lax eigenvector 
$|\psi_{\alpha,n}(t)|^2$ $\alpha=166$ 
(b,e) quasiparticle velocities computed by Eq.(\ref{qvel}) (solid red line) and 
corresponiding Lax eigenvalue
$\lambda_\alpha(t)$; (c,f) plots of $\lambda_\alpha(t)$ versus the time-averaged
velocity $\langle v_\alpha \rangle$. For comparison, in the Morse case 
(f) the average eigenvalue is reported.
In both cases, $N=200$ and initial conditions are sampled from 
a thermal GGE state of the Toda model with $\beta=1,\beta P=1$.
 }
 \label{fig1}
\end{figure*}

To start, it is convenient to write the equations of motion for the Toda lattice 
with the 
\textit{Flaschka variables} 
$a_j = \mathrm{e}^{-r_j/2},b_j =  p_j,$ (with $r_j \equiv q_{j+1} - q_j$) as
\begin{equation}
\label{eqmot}
\dot a_j = \tfrac{1}{2}a_j(p_j - p_{j+1}),\quad \dot p_j = a_{j-1}^2 - a_{j}^2,
\end{equation}
under periodic boundary conditions, 
$a_{-1} = a_N,p_{N+1} = p_1$.  
One then defines the \textit{Lax matrix}, $L=L^T$, $N\geq 3$, and its pair matrix $B=-B^T$ as 
\begin{eqnarray}
L_{i,j}=a_{i-1}\delta_{i-1,j} + p_i\delta_{i,j}+a_{i}\delta_{i+1,j} \\
B_{i,j}=\frac12 a_{i-1}\delta_{i-1,j}  - \frac12 a_i\delta_{i+1,j}
\end{eqnarray}
with $L_{1,N}=L_{N,1}=a_N$, $B_{1,N}=-B_{N,1}=a_N$.
It is well known \cite{babelon2003introduction} that Eqs. (\ref{eqmot}) 
can be recasted as 
$\dot L = [B,L]=BL -LB$. 

Consider the motion of a
perturbed Toda system in the general form \cite{bilman2016evolution}
\begin{equation}
\label{laxe} 
\dot  L = [B,L] + U
\end{equation}
where $U(a,p)$ is a $N\times N$ symmetric matrix with the same 
structure as $L$ (i.e. if $L_{i,j}=0$ then also $U_{i,j}=0$) whose elements are (nonlinear) functions of the Flashka variables.  

Consider the eigenvalue problem
\begin{equation}
\label{eigprob} 
L \ket{\alpha} = \lambda_\alpha \ket{\alpha}, \quad\alpha = 1,\dots,N.
\end{equation}
where $\lambda_\alpha$ and $\ket{\alpha}$ are the Lax eigenvalues and eigenvectors, whose components in the lattice basis $\ket{j}$ are 
$ \psi_{\alpha,j}=\braket{\alpha|j}$.

\section{Integrability} For Toda ($U=0$), $L(t)$ is isospectral i.e. $\lambda_\alpha$ 
do not change while eigenvectors are time-dependent and satisfy  
$\dot{\ket{\alpha}} = B\ket{\alpha}.$ 
At zero temperature the spectrum is 
$\lambda_\alpha=2\cos(2\pi \alpha/N)$ and eigenstates with $|\lambda_\alpha|>1$
are associated to solitons \cite{goldfriend2023effective}.
The physically relevant conserved charges are given by $\mathrm{tr}\big(L^n\big) = 
\sum_{j=1}^N (L^n)_{j,j} = \sum_{\alpha=1}^N  \lambda_\alpha^n$. 
The first three are the standard ones, namely the sum of the stretches $r_j$, momenta
$p_j$ and local energies $p_j^2 + a_j^2 + a_{j-1}^2$.

A thermodynamic state corresponds to a Generalized Gibbs Ensemble (GGE), 
with finite energy density fixed by the $N$
independent chemical potentials \cite{spohn2021hydrodynamic}. In this context, $L$ is a random matrix 
sampled from each GGE state. The simplest case would be the \textit{thermal} one where the assigned  
thermodynamic variables are the stretch $\ell = \langle r_j\rangle = -2\langle \log a_j\rangle$,  kinetic temperature $1/\beta=\langle p_j^2 \rangle$ and pressure $P=\langle a_j^2\rangle$ \cite{spohn2021hydrodynamic}.

Lax eigenvalues thus play a major role in the thermodynamics, 
as averages can be written as integrals over their empirical Density Of States (DOS)
$\rho(\lambda) = \lim_{N\to\infty}   \sum_{\alpha=1}^N \delta(\lambda - \lambda_\alpha)/N$ that
can be computed numerically by sampling the $L$ matrices and direct diagonalization or analytically via the thermodynamic Bethe Ansatz. 
The thermal DOS $\rho_{th}(\lambda)$ is of particular relevance:
in this case the $L_{i,j}$ are independent random variables
and the Lax spectrum can be sampled easily. Also, 
explicit analytical expressions of $\rho_{th}(\lambda)$  are 
available in some limit cases \cite{spohn2021hydrodynamic}.
Another important property is that  
the spectral gaps are proportional to Toda actions \cite{ferguson1982nonlinear}.

\section{Quasiparticles}
The quasiparticle concept is insightful to understand the dynamics
\cite{spohn2021hydrodynamic,politi2011heat,bulchandani2019kinetic,bulchandani2021quasiparticle}. For integrable systems quasiparticles move ballistically. Upon collisions they retain their velocity but undergo a spatial shift, the case termed \textit{interacting} in Ref. \cite{spohn2018interacting}. This results in an \textit{effective velocity}, which 
for Toda is solution of suitable integral equation \cite{spohn2020generalized}.
The idea that  these quasiparticles,  acquire an effective,   constant velocities
has been also rigorously established \cite{aggarwal2025effective}.  There is also 
an intriguing connection between such concepts and that 
of Anderson localization, rooted in the basically random nature of  Lax matrices
\cite{aggarwal2025asymptotic}.

To visualize this concept, one can define the 
quasiparticle position and velocity $x_\alpha,v_\alpha$
as the center of mass of the Lax eigenvector 
\cite{bulchandani2019kinetic}
\begin{equation}
v_\alpha=\dot x_\alpha = \frac{d}{dt}\sum_j \psi_{\alpha,j} q_j \psi_{\alpha,j} \equiv 
\sum_{ij} \psi_{\alpha,i} V_{i,j} \psi_{\alpha,j}
\label{qvel}
\end{equation}
where $V$ is symmetric,  tridiagonal with diagonal elements $p_j$ and upper
diagonal $\frac{1}{2} a_j \log a_j$ \cite{bulchandani2019kinetic}. 
At variance with $\lambda_\alpha$ this velocity is not constant due to the spatial shifts.

\section{Conservative perturbations of the Toda chain}
Upon multiplying both sides of Eq.(\ref{laxe}) by $L^{n-1}$, and using cyclic and linearity 
properties of the trace
\begin{equation}
\frac{1}{n} \frac{d}{dt} tr L^n  = tr(UL^{n-1}).
\end{equation}
So in general, the conservation laws of the eigenvalues are destroyed except
for the case $n=1$ for which $\sum_\alpha \lambda_\alpha$ is maintained in the class of momentum-conserving perturbations such that
$\mathrm{tr}\, U=0$. In the following, the energy-conserving case will be considered 
for which the  phase-space divergence,
$\sum_{ij,j\ge i} \frac{\partial \dot L_{ij}}{\partial L_{ij}}
\equiv \mathrm{div}\, \dot L =    \mathrm{div}\,U $, vanishes.
Thus, the condition $div U=0$ ensures that the perturbation is conservative.

\section{Examples}  In principle, for any Hamiltonian $H$
one can recast the equation of motion in the form (\ref{laxe}).
However, for a generic perturbation, $U$ may have a complicated
dependence on the $L_{i,j}$: for instance, terms like $r_j^p$ in $\Phi$
would yield entries proportional to $\log^{p-1} a_j$ \cite{bilman2016evolution}.  
It is thus useful to examine some simpler cases. 
The first is the \textit{Morse chain}
$\Phi(x)=(e^{-x/2}-\varepsilon)^2$ \cite{volkel1993quantum} then (up to a constant)
\begin{equation}
H_\mathrm{Morse}=H_\mathrm{Toda} +2\varepsilon \sum_j \mathrm{e}^{-\frac12 (q_{j+1} - q_j)},
\label{morse}
\end{equation}
corresponding to a perturbation matrix 
$U_{i,j} = 2\varepsilon (a_{j-1} - a_{j}) \, \delta_{ij}$
in Eq. (\ref{laxe}). The second example is the Toda model with non-uniform couplings 
$1+\varepsilon_j$ among neighbors, namely 
\begin{equation}
\label{coup}
H_\mathrm{C} = 
H_\mathrm{Toda} +\sum_{j}\varepsilon_j\mathrm{e}^{-(q_{j+1} - q_j)}
\end{equation}
for which 
$U_{i,j} = \left(\varepsilon_{j-1}a_{j-1}^2 - \varepsilon_{j}a_{j}^2\right)\delta_{ij}$.
For both examples, $U$ is diagonal (with $\mathrm{tr}\, U=0$) and its 
elements are, respectively, linear and quadratic in the $a_j$.
Thus Eqs.(\ref{laxe}) can be integrated
directly, which has some computational advantage since only evaluation of 
algebraic functions is required  
\footnote{Another class of example studied in the literature is 
the Toda with non-homogeneous
masses $m_j$ \cite{H99,garnier2003soliton}. It can be shown that also this case yields
a quadratic dependence of $U$ on the $a_j$, but  
$U$ is non-diagonal.  
}.

\section{Lax dynamics} 
The main idea is now to look at the time evolution of the $\lambda_\alpha$
that, under the effect of the perturbation $U$, are no longer constant 
\footnote{To the best of my knowledge, this point of has been considered only
very sporadically in the literature, see for example the case of  
the driven-dissipative Toda chain \cite{fesser1985chaos}
and Toda with a point-like defect \cite{geist1986chaos}. }.
For illustration,  let us 
compare simulations of the Toda and Morse chains.
In agreement with intuition, the space-time evolution of a Lax eigenvector 
$|\psi_{\alpha,n}(t)|^2$ [Fig. \ref{fig1}(a)]     
looks soliton-like in the integrable case, propagating ballistically with random 
space shifts, see Fig. \ref{fig1}(b) where the quasi-particle velocity 
$v_\alpha(t)$ is reported along with the corresponding (constant) $\lambda_\alpha$.  
Plotting $\lambda_\alpha$ versus the time-averaged velocity 
$\langle v_\alpha\rangle$ [Fig. \ref{fig1}(c)] confirms 
an approximate correspondence between the two quantities, indicating 
that one can be used as a proxy of the other.
Also, they become almost exactly proportional for soliton
modes located closer to the Lax spectrum band edges.   

Remarkably, the above picture remains similar also in presence 
of the perturbation. The main difference is that the almost-ballistic
propagation is interrupted by huge scattering shifts, where the 
center of mass of the eigenvector very rapidly "jumps" to another
site [Fig. \ref{fig1}(d)]. Also,   Fig. \ref{fig1}(e,f) confirms
that there is a close correspondence between the Lax eigenvalues 
and the effective velocities
\footnote{Strictly speaking, expression (\ref{qvel}) holds only for 
$U=0$: in presence of the perturbation there is a correction 
due to additional scattering that we neglect for the sake of
simplicity, since we just perform a qualitative comparison here.}
. 

\begin{figure}
\centering \includegraphics[width=0.45\textwidth]{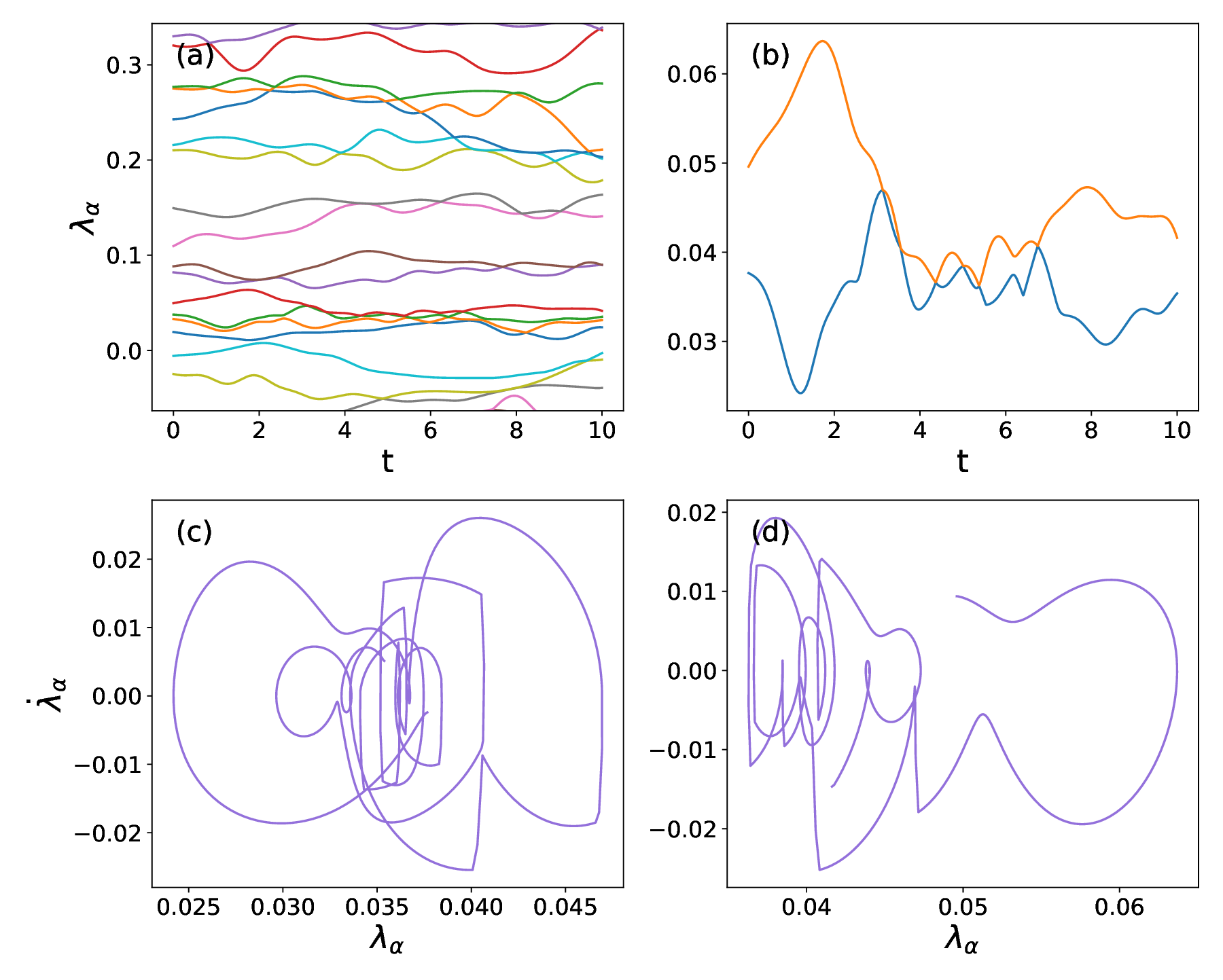}
\caption{Lax dynamics for the Morse chain, Eq. (\ref{morse}): 
time evolution of (a) a subset
of eigenvalues $\lambda_\alpha(t)$ and (b) a couple of neighboring
ones $\lambda_\alpha(t),\lambda_{\alpha+1}(t)$ illustrating the strong repulsion
that yields almost elastic collisions ;
(c,d): 
phase portraits 
($\lambda_\alpha,\dot{\lambda}_\alpha$) for $\alpha=102,103$.  The chain is initialized with
random initial conditions sampled from the Toda thermal GGE state with 
$\beta=P=1$, $N=200$, $\varepsilon=0.1$.}
\label{fig2}
\end{figure}

The eigenvalues trajectories for the Morse chain [Fig. \ref{fig2}(a,b)]
manifestly behave as a one-dimensional "gas" of particles. In the 
simulation, the gas remains confined in a bounded region, with no 
escape, at least on the considered time-scale. 
A remarkable feature is that two neighboring eigenvalues
undergo almost elastic collisions, in which they 
approximately exchange their values, as clearly seen in Fig. \ref{fig2}(b).
Accordingly, this induces relatively large changes in the 
velocity $v_\alpha$ which accounts for the large scattering shifts 
observed in Fig. \ref{fig1}(d). In the phase portraits this
yields abrupt changes in $\dot \lambda_\alpha$, 
Fig. \ref{fig2}(c,d). 

For an analytical formulation of Lax dynamics, one proceeds by  
computing $d(L\ket{\alpha})/dt$ from Eq. (\ref{eigprob}) and 
using Eq. (\ref{laxe}), yielding 
\begin{equation}
U \ket{\alpha} +(L-\lambda_\alpha)\big(\dot {\ket{\alpha}} -B \ket{\alpha}\big)
=\dot \lambda_\alpha\ket{\alpha}.
\end{equation}
Upon left-multipying by $\bra{\beta}$, 
using $\braket{\alpha|\beta}=\delta_{\alpha,\beta}$, 
and letting $\braket{\alpha |U|\beta}\equiv U_{\alpha\beta}$ one obtains 
from the diagonal elements that $\dot \lambda_\alpha= U_{\alpha\alpha}$
and from the non-diagonal elements the evolution equation
for the eigenvectors
\begin{equation}
\ket{\dot\alpha}= B \ket{\alpha} \,+\, \sum_{\beta\neq \alpha} 
\frac{ U_{\alpha\beta}}{\lambda_\alpha-\lambda_\beta} \ket{\beta}
\label{alfadot}
\end{equation} 
which reduces to the unperturbed Toda evolution for $U=0$, and 
is reminiscent of a quantum Hamiltonian with hopping terms 
induced by the perturbation.

\begin{figure}
\includegraphics[scale=0.35]{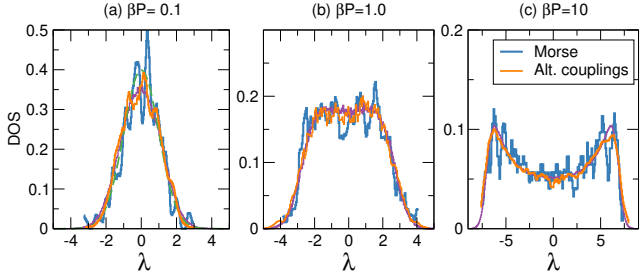}
\caption{The DOS $\rho(\lambda)$ obtained from the Lax dynamics of the 
Morse (blue) and the Toda chain with  alternating (staggered) coupling 
$\varepsilon_j = (-1)^j \varepsilon$ (orange lines), ($N=200$, $\varepsilon=0.1$), 
starting with initial thermal GGE
initial conditions with three different $\beta P$, $\beta=1$. 
Magenta lines are the thermal DOS $\rho_{th}$ for the unperturbed Toda chain, 
obtained by sampling and diagonalizing the equilibrium Lax matrix.
The dashed green line in (a) is the approximate Gaussian DOS expected 
predicted in the limit of small $\beta P$ \cite{spohn2021hydrodynamic}.}
\label{fig3}
\end{figure}

Finally, one computes $\dot U_{\alpha\beta}$ and, 
using Eq.(\ref{alfadot}), 
\begin{eqnarray}
\label{lambdadot}
&&\frac{d\lambda_\alpha}{dt} = U_{\alpha\alpha}\equiv \pi_\alpha \nonumber\\
&&\frac{d\pi_{\alpha}}{dt} =
F_{\alpha\alpha}
+2\sum_{\beta\neq \alpha} 
\frac{U_{\alpha\beta}^2}{\lambda_\alpha-\lambda_\beta} \\
&&\frac{dU_{\alpha\beta}}{dt}  = F_{\alpha\beta}
+ U_{\alpha\beta} \frac{U_{\beta\beta}-U_{\alpha\alpha}}{\lambda_\alpha-\lambda_\beta}
\nonumber+\\
&&+\sum_{\gamma\neq \alpha,\beta} U_{\alpha\gamma}U_{\gamma\beta}
\left[\frac{1}{\lambda_\alpha-\lambda_\gamma}  + 
\frac{1}{\lambda_\beta-\lambda_\gamma}\right] \qquad(\alpha\neq\beta)
\nonumber
\end{eqnarray}
where $F_{\alpha\beta}\equiv  \braket{\alpha |\left(\dot U + [U,B]\right)|\beta}$.
Eqs. (\ref{lambdadot}) are exact and hold
\textit{whatever the form of} $U$.  
Also, using the invariance of the trace 
with respect to change of basis,  if momentum conservation holds for 
also in presence of perturbation  $tr U=0$,  the
the "momentum", $\sum_\alpha \pi_\alpha$ is conserved too. 

If $F_{\alpha\beta}$ would vanish, Eqs. (\ref{lambdadot}) would be \textit{identical}
to the so-called Pechukas-Yukawa (PY) equations ruling the evolution of the 
eigenenergies of a quantum Hamiltonian of the form $\mathcal H=\mathcal H_0+t\mathcal H_1$, as a function of the 
fictitious "time" parameter
$t$ \cite{pechukas1983distribution,yukawa1985new}. Also,  
Eq. (\ref{alfadot}) would formally correspond to those for 
quantum eigenstates \cite{nakamura1986complete}. 
The PY equations are closed and 
define a $2N+N(N-1)/2$-dimensional dynamical system ($U$ is symmetric)
with generalized coordinates $(\lambda_\alpha,\pi_\alpha,U_{\alpha\beta})$.
It has the remarkable features to be both Hamiltonian  
and fully integrable. Indeed, the change of variable 
$V_{\alpha\beta} ={U_{\alpha\beta}}{(\lambda_\alpha-\lambda_\beta)}$ ($\alpha\neq\beta$) 
transforms the PY equations 
in a generalized version of the famous Calogero-Moser model
\cite{nakamura1986complete,yukawa1986lax}. 

Yet, the presence of the terms $F_{\alpha\beta}$  hinders an
exact analysis as in the PY case: indeed, Eqs. (\ref{lambdadot}) are not 
closed and, in general, are also explicitly time-dependent
\footnote{Indeed also in the quantum problem, a different (nonlinear) dependence of the
Hamiltonian $\mathcal H(t)$ on the parameter $t$ may yield additional non-autonomous terms, that
can be handled analytically only is specific cases, see e.g. Ref.\cite{haake1990classical}}.
Despite those difficulties, some interesting physical consequences
can be envisaged. 

First, ignoring for the moment the terms $F_{\alpha\alpha}$,  
the first two of Eqs.(\ref{lambdadot}) describe
a one-dimensional Dyson-Coulomb gas   
coupled through the fluctuating "charges" $U_{\alpha\beta}^2$ 
provided by the remaining $N(N-1)/2$ degrees of freedom that act as a "heat bath" 
\cite{yang1991molecular}.  Indeed, for a finite number
of levels, the DOS is approximatively given by the the invariant measure of 
the Coulomb gas under an external quadratic potential \cite{yukawa1985new}. 
On the other hand, for the Toda chain, such measure coincides also with the 
thermal DOS $\rho_{th}$ \cite{spohn2021hydrodynamic}. 
Indeed, based on the above heuristic consideration, one may surmise that the Lax dynamics 
naturally provides a general thermalization pathway towards the thermal DOS for 
any perturbation in the class of the above examples.  
In other words,  the fluctuations of the charges provide 
the  chaos (noise)  source needed to thermalize the Dyson gas. 
It may be argued, that the generic mechanism leading to thermalization 
is provided by level repulsion and quasi-elastic scattering
seen in Fig.\ref{fig2}.

To support this idea, the data in Fig. \ref{fig3}, show the DOS $\rho(\lambda)$
for models (\ref{morse}) and (\ref{coup}) are basically independent of the 
choice of $U$.  Moreover,  for all the simulations considered here,  the 
equilibrium DOS 
$\rho(\lambda)\approx\rho_{th}(\lambda)$, within statistical fluctuations.
So,  the thermal DOS of the unperturbed
Toda accurately describes the DOS of the non-integrable models.

As a final remark about integrability, a numerical simulation of the Lax dynamics in simplest 
case $N=3$ indicate that the trajectories are compatible with quasiperiodic 
motion on invariant tori, a hint that some form of integrability may
occur also here. This certainly deserve a closer mathematical analysis. 
  
\section{Conclusions}  
Lax dynamics is a novel and insightful approach
to describe the effect of perturbations on a many-body integrable system
at finite energy density. Its implementation is computationally straightforward, and it has the potential to be extended to other systems, such as perturbations of integrable discretizations of the nonlinear Schrödinger equation \cite{salerno1992quantum}, among others. It offers a physically appealing interpretation
in terms of quasi-particle collisions as avoided eigenvalue crossings. Furthermore, in the case of weak perturbations, Lax dynamics is expected to evolve on a slower timescale compared to the natural timescale of the Flashka variables. Therefore, it could be an effective approach for studying the slow evolution of Toda actions in the context of thermalization problems \cite{benenti2023non,goldfriend2019equilibration}.
Finally, the analogy with the PY gas is highly suggestive and warrants a more detailed investigation. Equations (\ref{lambdadot}) may allow for an effective, reduced dynamical description of the relevant quantities.
The 'universal' evolution described by PY dynamics is one of the arguments used to justify the universal spectral statistics of quantum chaos.
Could a similar consideration also apply in the present context?
This could be one of the many possible research routes originating from the present work.

\acknowledgments
 Support is acknowledged from
the PRIN 2022 project \textit{Breakdown of ergodicity in classical and quantum
many-body systems }(BECQuMB) Grant No. 20222BHC9Z
CUP G53C24000680006 funded by the European Union—
NextGenerationEU, M4 C2 1.1. I thank L. Cugliandolo, 
S.Iubini, A. Politi and S. Ruffo for
useful discussions.

\textit{Data availability statement:} The data that support the
ﬁndings of this study are available upon reasonable request
from the author.

\bibliographystyle{eplbib}
\bibliography{quasint}

\begin{thebibliography}{10}
\expandafter\ifx\csname url\endcsname\relax\def\url#1{\texttt{#1}}\fi

\bibitem{bastianello2021hydrodynamics}
\Name{Bastianello A., De~Luca A. \and Vasseur R.} \REVIEW{Journal of
  Statistical Mechanics: Theory and Experiment}{2021}{2021}{114003}.

\bibitem{doyon2025generalized}
\Name{Doyon B., Gopalakrishnan S., M{\o}ller F., Schmiedmayer J. \and Vasseur
  R.} \REVIEW{Physical Review X}{15}{2025}{010501}.

\bibitem{toda2012theory}
\Name{Toda M.} \Book{Theory of nonlinear lattices} Vol.~20 (Springer Science \&
  Business Media) 2012.

\bibitem{henon1974integrals}
\Name{Hénon M.} \REVIEW{Phys. Rev. B}{9}{1974}{1921}.

\bibitem{Flaschka1974}
\Name{Flaschka H.} \REVIEW{Physical Review B}{9}{1974}{1924}.

\bibitem{theodorakopoulos1984finite}
\Name{Theodorakopoulos N.} \REVIEW{Physical review letters}{53}{1984}{871}.

\bibitem{cuccoli1994thermodynamics}
\Name{Cuccoli A., Livi R., Spicci M., Tognetti V. \and Vaia R.}
  \REVIEW{International Journal of Modern Physics B}{8}{1994}{2391}.

\bibitem{doyon2019generalized}
\Name{Doyon B.} \REVIEW{Journal of Mathematical Physics}{60}{2019}{073302}.

\bibitem{spohn2020generalized}
\Name{Spohn H.} \REVIEW{Journal of Statistical Physics}{180}{2020}{4}.

\bibitem{spohn2021hydrodynamic}
\Name{Spohn H.} \REVIEW{arXiv preprint arXiv:2101.06528}{}{2021}{}.

\bibitem{baldovin2021statistical}
\Name{Baldovin M., Vulpiani A. \and Gradenigo G.} \REVIEW{Journal of
  Statistical Physics}{183}{2021}{41}.

\bibitem{dauxois2006physics}
\Name{Dauxois T. \and Peyrard M.} \Book{Physics of solitons} (Cambridge
  University Press) 2006.

\bibitem{kundu2016equilibrium}
\Name{Kundu A. \and Dhar A.} \REVIEW{Physical Review E}{94}{2016}{062130}.

\bibitem{mazzuca2023equilibrium}
\Name{Mazzuca G., Grava T., Kriecherbauer T., McLaughlin K. T.-R., Mendl C.~B.
  \and Spohn H.} \REVIEW{Journal of Statistical Physics}{190}{2023}{149}.

\bibitem{gallavotti2007fermi}
\Name{Gallavotti G.} (Editor) \Book{The Fermi-Pasta-Ulam Problem: A Status
  Report} Vol. 728 of \emph{Lecture Notes in Physics} (Springer) 2008.

\bibitem{benenti2023non}
\Name{Benenti G., Donadio D., Lepri S. \and Livi R.} \REVIEW{La Rivista del
  Nuovo Cimento}{46}{2023}{105}.

\bibitem{benettin2013fermi}
\Name{Benettin G., Christodoulidi H. \and Ponno A.} \REVIEW{Journal of
  Statistical Physics}{152}{2013}{195}.

\bibitem{benettin2021understanding}
\Name{Benettin G. \and Ponno A.} \REVIEW{Mathematics in
  Engineering}{3}{2021}{1}.

\bibitem{goldfriend2019equilibration}
\Name{Goldfriend T. \and Kurchan J.} \REVIEW{Physical Review
  E}{99}{2019}{022146}.

\bibitem{benettin2023role}
\Name{Benettin G., Orsatti G. \and Ponno A.} \REVIEW{Journal of Statistical
  Physics}{190}{2023}{131}.

\bibitem{chen2014nonintegrability}
\Name{Chen S., Wang J., Casati G. \and Benenti G.} \REVIEW{Physical Review
  E}{90}{2014}{032134}.

\bibitem{DiCintio2018}
\Name{{Di Cintio} P., Iubini S., Lepri S. \and Livi R.} \REVIEW{Chaos, Solitons
  \& Fractals}{117}{2018}{249}.

\bibitem{dhar2019transport}
\Name{Dhar A., Kundu A., Lebowitz J.~L. \and Scaramazza J.~A.} \REVIEW{Journal
  of Statistical Physics}{175}{2019}{1298}.

\bibitem{lepri2020too}
\Name{Lepri S., Livi R. \and Politi A.} \REVIEW{Physical Review
  Letters}{125}{2020}{040604}.

\bibitem{babelon2003introduction}
\Name{Babelon O., Bernard D. \and Talon M.} \Book{Introduction to classical
  integrable systems} (Cambridge University Press) 2003.

\bibitem{bilman2016evolution}
\Name{Bilman D. \and Nenciu I.} \REVIEW{Physica D: Nonlinear
  Phenomena}{330}{2016}{1}.

\bibitem{goldfriend2023effective}
\Name{Goldfriend T.} \REVIEW{Journal of Statistical Physics}{190}{2023}{70}.

\bibitem{ferguson1982nonlinear}
\Name{Ferguson~Jr W., Flaschka H. \and McLaughlin D.} \REVIEW{Journal of
  computational physics}{45}{1982}{157}.

\bibitem{politi2011heat}
\Name{Politi A.} \REVIEW{Journal of Statistical Mechanics: Theory and
  Experiment}{2011}{2011}{P03028}.

\bibitem{bulchandani2019kinetic}
\Name{Bulchandani V.~B., Cao X. \and Moore J.~E.} \REVIEW{Journal of Physics A:
  Mathematical and Theoretical}{52}{2019}{33LT01}.

\bibitem{bulchandani2021quasiparticle}
\Name{Bulchandani V.~B., Kulkarni M., Moore J.~E. \and Cao X.} \REVIEW{Journal
  of Physics A: Mathematical and Theoretical}{54}{2021}{474001}.

\bibitem{spohn2018interacting}
\Name{Spohn H.} \REVIEW{Journal of Mathematical Physics}{59}{2018}{}.

\bibitem{aggarwal2025effective}
\Name{Aggarwal A.} \REVIEW{arXiv:2503.11407}{}{2025}{}.

\bibitem{aggarwal2025asymptotic}
\Name{Aggarwal A.} \REVIEW{arXiv:2503.08018}{}{2025}{}.

\bibitem{volkel1993quantum}
\Name{V{\"o}lkel A., Cuccoli A., Spicci M. \and Tognetti V.} \REVIEW{Physics
  Letters A}{182}{1993}{60}.

\bibitem{H99}
\Name{Hatano T.} \REVIEW{Phys. Rev. E}{59}{1999}{R1}.

\bibitem{garnier2003soliton}
\Name{Garnier J. \and Abdullaev F.~K.} \REVIEW{Physical Review
  E}{67}{2003}{026609}.

\bibitem{fesser1985chaos}
\Name{Fesser K., McLaughlin D., Bishop A. \and Holian B.} \REVIEW{Physical
  Review A}{31}{1985}{2728}.

\bibitem{geist1986chaos}
\Name{Geist K. \and Lauterborn W.} \REVIEW{Physica D: Nonlinear
  Phenomena}{23}{1986}{374}.

\bibitem{pechukas1983distribution}
\Name{Pechukas P.} \REVIEW{Physical review letters}{51}{1983}{943}.

\bibitem{yukawa1985new}
\Name{Yukawa T.} \REVIEW{Physical review letters}{54}{1985}{1883}.

\bibitem{nakamura1986complete}
\Name{Nakamura K. \and Lakshmanan M.} \REVIEW{Physical review
  letters}{57}{1986}{1661}.

\bibitem{yukawa1986lax}
\Name{Yukawa T.} \REVIEW{Physics Letters A}{116}{1986}{227}.

\bibitem{haake1990classical}
\Name{Haake F. \and Lenz G.} \REVIEW{Europhysics Letters}{13}{1990}{577}.

\bibitem{yang1991molecular}
\Name{Yang X. \and Burgd{\"o}rfer J.} \REVIEW{Physical review
  letters}{66}{1991}{982}.

\bibitem{salerno1992quantum}
\Name{Salerno M.} \REVIEW{Phys. Rev. A}{46}{1992}{6856}.
\newline\url{https://link.aps.org/doi/10.1103/PhysRevA.46.6856}

\end{thebibliography}

\end{document}